\documentclass[twocolumn]{aastex62}

\usepackage{graphicx}	
\usepackage{amsmath}	
\usepackage{amssymb}	
\usepackage{bm}

    \newcommand\rd{{\mathrm{d}}}
    
    \newcommand\Ok{{\Omega_{k}}}
    
    \DeclareMathOperator{\Cov}{{\mathrm{Cov}}}
    \DeclareMathOperator{\Var}{{\mathrm{Var}}}

    \makeatletter
    
    \newcommand{\Rmnum}[1]{\expandafter\@slowromancap\romannumeral #1@}
    \makeatother
    
\received{xxx}
\revised{xxx}
\accepted{xxx}
\submitjournal{ApJ}

\shorttitle{Measuring spatial curvature with HIIGx and $H(z)$}
\shortauthors{C.-Z. Ruan, F. Melia, Y. Chen and T.-J. Zhang}

\begin{document}

\title{Using spatial curvature with HII galaxies and cosmic chronometers to explore the tension in $H_0$}

\correspondingauthor{Tong-Jie Zhang}
\email{tjzhang@bnu.edu.cn}

\author{Cheng-Zong Ruan}
\affiliation{Department of Astronomy, Beijing Normal University, \\
Beijing 100875, China}

\author{Fulvio Melia}
\affiliation{Department of Physics, The Applied Math Program, and Department of Astronomy, \\
The University of Arizona, AZ 85721, USA; fmelia@email.arizona.edu}

\author{Yu Chen}
\affiliation{Department of Astronomy, Beijing Normal University, \\
Beijing 100875, China}

\author{Tong-Jie Zhang}
\affiliation{Department of Astronomy, Beijing Normal University, \\
Beijing 100875, China}
\affiliation{Institute for Astronomical Science, Dezhou University, \\
Dezhou 253023, China}



\begin{abstract}

We present a model-independent measurement of spatial curvature $\Omega_{k}$ in the Friedmann-Lema\^itre-Robertson-Walker (FLRW) universe, based on observations of the Hubble parameter $H(z)$ using cosmic chronometers, and a Gaussian Process (GP) reconstruction of the HII galaxy Hubble diagram. 
We show that the imposition of spatial flatness (i.e., $\Omega_k=0$) easily distinguishes between the Hubble constant measured with {\it Planck} and that based on the local distance ladder. We find an optimized curvature parameter $\Omega_{k} = -0.120^{+0.168}_{-0.147}$ when using the former (i.e., $H_0=67.66\pm0.42 \, \mathrm{km}\,\mathrm{s}^{-1} \,\mathrm{Mpc}^{-1}$), and $\Omega_{k} = -0.298^{+0.122}_{-0.088}$ for the latter ($H_0=73.24\pm 1.74 \,\mathrm{km}\,\mathrm{s}^{-1} \,\mathrm{Mpc}^{-1}$). The quoted uncertainties are extracted by Monte Carlo sampling, taking into consideration the covariances between the function and its derivative reconstructed by GP. These data therefore reveal that the condition of spatial flatness favours the {\it Planck} measurement, while ruling out the locally inferred Hubble constant as a true measure of the large-scale cosmic expansion rate at a confidence level of $\sim 3\sigma$. 
\end{abstract}



\section{Introduction}
\label{sec:intro}
One of the fundamental assumptions in modern cosmology is that, on large scales, the Universe is described by the homogeneous and isotropic Friedmann-Lema\^itre-Robertson-Walker (FLRW) metric. The symmetries of this spacetime reduce the ten independent components of the metric tensor to a single function of time---the scale factor $a(t)$, and a constant---the spatial curvature parameter $k$, which may take on the values $-1$ (for an open Universe), $+1$ (closed) or $0$ (flat). The constant $k$ is often absorbed into the so-called curvature density parameter, $\Omega_{k} \equiv -kc^2 / (a_0 H_0)^2$, where $c$ is the speed of light, and $a_0 \equiv a(t_0)$ and the Hubble parameter $H_0\equiv H(t_0)$ take on their respective values at time $t_0$ (i.e., today). Thus, the Universe is open if $\Omega_{k} > 0$, spatially flat if $\Omega_{k} = 0$ and closed if $\Omega_{k} < 0$. Knowing which of these three possibilities describes the Universe is crucial for a complete understanding of its evolution and the nature of dark energy. A significant deviation from zero spatial curvature would have a profound impact on the underlying physics and the inflation paradigm, in part because one of the roles attributed to the inflaton field is that of rapidly expanding the Universe to asymptotic flatness, regardless of whether or not it was flat to begin with.

Current cosmological observations strongly favour a flat (or nearly flat) Universe, e.g., based on combined {\it Planck} and baryon acoustic oscillation (BAO) measurements, that suggest $\Omega_{k} = 0.0007 \pm 0.0019$ at the $68\%$ confidence level  \citep{2018arXiv180706209P}.\footnote{The {\it Planck} 2018 cosmic microwave background (CMB) temperature and polarization power spectra data singly favour a mildly closed Universe, i.e., $\Omega_{k} = -0.044^{+0.018}_{-0.015}$  \citep{2018arXiv180706209P}. Other studies have found that the {\it Planck} 2015 CMB anisotropy data also favour a mildly closed universe \citep{2018arXiv180305522P,2018arXiv180100213P}.} These constraints, however, are based on the pre-assumption of a particular cosmological model, such as $\Lambda$CDM. Since the curvature parameter is a purely geometric quantity, however, it should be possible to measure or constrain the value of $\Omega_{k}$ from the data using a model-independent method. For a non-exhaustive set of references on this topic, see \citet{2006ApJ...637..598B,2006PhRvD..73b3503K,2007JCAP...08..011C,2012AJ....143..120O,2014ApJ...789L..15L,PhysRevLett.115.101301,2016PhRvD..93d3517C,2016ApJ...828...85Y,2016ApJ...833..240L,2016PhRvD..93d3014L,2017ApJ...838..160W,2017ApJ...834...75X,2018ApJ...854..146L,2018JCAP...03..041D}, and \citet{2018arXiv180609781W}. A typical curvature measurement methodology is based on the distance sum rule in the FLRW metric using strong lensing \citep{2006ApJ...637..598B,PhysRevLett.115.101301}, that provides the angular diameter distance between the observer and lens, the observer and source, and the lens and source.

In this paper, we follow a new, model-independent methodology, that combines the observed Hubble parameter $H(z)$ with an independent measurement of the luminosity distance $d_L(z)$ \citep{2007JCAP...08..011C}. With this approach, one needs to have a continuous realization of the distance $d_L(z)$ and its derivative $d_L^\prime(z)$ with respect to redshift. A model-independent smoothing technique, based on the use of Gaussian processes (GP), can provide these quantities together with their respective uncertainties and covariances (see, e.g., \citet{2012JCAP...06..036S,yennapureddy2017reconstruction}). Using GP reconstruction, one can calculate a continuous luminosity distance and its derivative using HII galaxies (HIIGx) and Giant extragalactic HII regions (GEHR) as standard candles (\citet{terlevich1981dynamics,terlevich2015road,2012MNRAS.425L..56C,doi:10.1093/mnras/stu987,2016MNRAS.463.1144W}; and other references cited therein). Then, the luminosity distance $d_L(z)$ may be transformed into the curvature-dependent Hubble parameter $H (z;\Omega_{k})$ according to geometric relations derived from the FLRW metric. Finally, by carrying out $\chi^2$ minimization on the observed differences between $H(z)$ and $H(z;\Omega_{k})$, one may thereby optimize the value of  $\Omega_k$ in a model-independent way.

There are several possible applications of this approach that will be explored in subsequent papers. Here, we apply this method to one of the most timely and important problems emerging from the latest cosmological data---the non-ignorable tension between the value of $H_0$ measured with the local distance ladder (e.g., \citealt{2016ApJ...826...56R})---consistently yielding a value $\sim 73\, \mathrm{km}\,\mathrm{s}^{-1} \,\mathrm{Mpc}^{-1}$ and an impressively small error of $\sim 2\text{-}3\%$---and the value measured with {\it Planck}, based on the fluctuation spectrum of the cosmic microwave background (CMB) \citep{2018arXiv180706209P}, i.e., $67.66\pm 0.42\, \mathrm{km}\,\mathrm{s}^{-1} \,\mathrm{Mpc}^{-1}$. These two measurements of $H_0$ are discrepant at a level {\it exceeding} $3\sigma$. Measurements of the spatial curvature parameter $\Omega_k$ are often invoked to test inflationary theory, given that a principal role of the inflaton field is to drive the universal expansion to asymptotic flatness. In this paper, we reverse this procedure by instead presuming that the Universe is flat and using the $H(z)$ and HII galaxy observations  to then examine which of these two values of $H_0$ is more consistent with this assumption.

We first briefly summarize the methodology of measuring $\Omega_{k}$ using HII galaxies and cosmic chronometers in \S~\ref{sec:metho}. We then describe the relevant data sets in \S~\ref{sec:data}, and present the results of our analysis in \S~\ref{sec:res}. We then discuss these results in \S~\ref{sec:conc}, where we conclude that this test strongly favours the {\it Planck} value as the true representation of the expansion rate on large scales.

\section{Methodology}
\label{sec:metho}
\subsection{Luminosity Distance and distance modulus}

The hydrogen gas ionized by massive star clusters in HII galaxies emits prominent Balmer lines in $\mathrm{H}\alpha$ and $\mathrm{H}\beta$  \citep{terlevich1981dynamics,kunth2000most}.
The luminosity $L(\mathrm{H}\beta)$ in $\mathrm{H}\beta$ in these systems is strongly correlated with the ionized gas velocity dispersion $\sigma_v$ of the ionized gas \citep{terlevich1981dynamics}, (presumably) because both the intensity of ionizing radiation and $\sigma_v$ increase with the starbust 
mass \citep{siegel2005towards}. 
The relatively small scatter in the relationship between $L(\mathrm{H}\beta)$ and $\sigma_v$ allows these galaxies and local HII regions to be used as standard candles \citep{terlevich2015road,2016MNRAS.463.1144W,yennapureddy2017reconstruction,2018MNRAS.474.4507L}. 
The emission-line luminosity versus ionized gas velocity dispersion correlation can be approximated as \citep{2012MNRAS.425L..56C}
\begin{align}
\log \left[ \frac{L (\mathrm{H} \beta)}{\mathrm{erg}\,\mathrm{s}^{-1}} \right] = \alpha\, \log \left[ \frac{\sigma_v (\mathrm{H} \beta)}{\mathrm{km} \, \mathrm{s}^{-1}} \right] + \kappa \ , \label{equ:Lsigmarel}
\end{align}
where $\alpha$ and $\kappa$ are constants.  
With this $L \text{-} \sigma_v$ relation, and the luminosity distance of an HIIGx, 
\begin{equation}
    d_L = \left[ \frac{L (\mathrm{H} \beta)}{4 \pi F (\mathrm{H} \beta)} \right]^{1/2}\;,
\end{equation}
we write
\begin{align} 
    \log \left( \frac{d_L}{\mathrm{Mpc}} \right) &= 0.5 \left[ \alpha \log \left( \frac{\sigma_v (\mathrm{H}\beta)}{\mathrm{km}\,\mathrm{s}^{-1}} \right) - \right.\nonumber\\
&\quad\left.\log \left( \frac{F (\mathrm{H} \beta)}{\mathrm{erg}\,\mathrm{s}^{-1} \,
\mathrm{cm}^{-2}} \right) \right] + 0.5 \kappa - 25.04 \ ,
\end{align}
where $F (\mathrm{H} \beta)$ is the reddening corrected $\mathrm{H} \beta$ flux.

Note that the luminosity distance $d_L \equiv \sqrt{ L / (4\pi F)}$ is conventionally defined by the bolometric 
luminosity and flux, whereas here we approximate this relation with the luminosity $L (\mathrm{H} \beta)$ and
flux $F(\mathrm{H} \beta)$ pertaining to the $\mathrm{H} \beta$ line. When selecting HII galaxies from spectroscopic 
surveys, the most import criteria are (i) the identification of the largest equivalent width (EW) in the galaxies' 
emission lines, with $\mathrm{EW} (\mathrm{H} \beta) > 50 \,\text{\AA}$ or $\mathrm{EW} (\mathrm{H} \alpha) > 200 \,
\text{\AA}$ in their rest frame and (ii) that the emission regions are extremely compact. The lower limits guarantee 
that the selected HIIGx are comprised of systems in which the luminosity is dominated by single and very young 
starbursts (less than $5 \,\mathrm{Myr}$ in age) \citep{terlevich2015road}. The bolometric flux of the HIIGx may
thereby be regarded as constituting principally the $\mathrm{H} \beta$ line.

The two `nuisance' parameters $\alpha$ and $\kappa$ (or its Hubble-free replacement, $\delta$; see 
definition below) in principle need to be optimized simultaneously with those of the cosmological model. 
\citet{2016MNRAS.463.1144W} have found, however, that their values deviate by at most only a tiny fraction of their 
$1\sigma$ errors, regardless of which model is adopted. This is the important step that allows us to use the HII galaxy 
Hubble diagram in a model-independent way. For example, these authors found that $\alpha=4.86^{+0.08}_{-0.07}$ and 
$\delta = 32.38^{+0.29}_{-0.29}$ in the $R_{\rm h}=ct$ cosmology \citep{2012MNRAS.419.2579M}, while 
$\alpha = 4.89^{+0.09}_{-0.09}$ and $\delta = 32.49^{+0.35}_{-0.35}$ in $\Lambda$CDM. Such small differences fall 
well within the observational uncertainty and, following \citet{melia2018model}, we therefore simply adopt a 
reasonable representation of the average value for these parameters, i.e., $\alpha = 4.87^{+0.11}_{-0.08}$ and 
$\delta = 32.42^{+0.42}_{-0.33}$. We emphasize, however, that this relative model-independence
of $\alpha$ and $\delta$ is based on the best-fit analysis using $R_{\rm h}=ct$ and various versions of 
$\Lambda$CDM \citep{2016MNRAS.463.1144W}. The caveat here is that if one chooses an even more exotic model that 
differs from $\Lambda$CDM and $R_{\mathrm{h}}=ct$ by greater amounts, these parameters may themselves vary more
strongly than we assume here.

The Hubble-free parameter $\delta$ is defined as follows: 
\begin{align}
    \delta \equiv -2.5 \kappa - 5 \log \left( \frac{H_0}{\mathrm{km}\,\mathrm{s}^{-1}\, \mathrm{Mpc}^{-1}} \right) + 125.2\;,
\end{align}
with which one may express the dimensionless luminosity distance $(H_0\,d_L)/c$ as
\begin{align}
    \left(\frac{H_0}{c}\right) d_L(z) = \frac{10^{\eta(z)/5}}{ c / (\mathrm{km}\,\mathrm{s}^{-1}) } \;, \label{equ:dLnu}
\end{align}
where 
\begin{align}
    \eta  & = -\delta + 2.5\left(\alpha\, \log \left[ \frac{\sigma_v (\mathrm{H}\beta)}{\mathrm{km}\, \mathrm{s}^{-1}} \right] - \log \left[ \frac{F(\mathrm{H}\beta)}{\mathrm{erg}\, \mathrm{s}^{-1} \, \mathrm{cm}^{-2}} \right] \right) \ , \label{equ:nudef}
\end{align}
and the speed of light $c$ is $299792.458 \,\mathrm{km}\,\mathrm{s}^{-1}$.
Equation~\eqref{equ:nudef} is an \textit{approximation} for the distance modulus $\mu$ under the 
assumption of the $L\text{-} \sigma_v$ relation in Equation~\eqref{equ:Lsigmarel}. In addition, $\eta(z)$
defined in Equation~\eqref{equ:dLnu}  differs by a constant from 
\begin{align}
    \mu &\equiv 5 \log \left( \frac{d_L}{\mathrm{Mpc}} \right) + 25 \ , \\
\intertext{such that}
    \mu - \eta &= -5 \log \left( \frac{H_0}{\mathrm{km} \, \mathrm{s}^{-1} \, \mathrm{Mpc}^{-1}} \right) + 25.2 \ .
\end{align}

Given the flux and gas velocity dispersion (along with their uncertainties) of HIIGx and GEHR, one can get the `shifted' distance modulus $\eta(z)$ using Equation \eqref{equ:nudef}. And for each measurement of $H(z_j)$ using cosmic chronometers at redshift $z_j$, we use a model-independent GP reconstruction to get the corresponding $\eta(z_j)$ and ${\eta}^\prime(z_j)$ as well as their uncertainties and covariances, where the derivative is defined by 
\begin{align}
    {\eta}^\prime \equiv \frac{d \eta}{d \log_{10} z} \ .
\end{align}

The systematic uncertainties of the HII-galaxy probe, based on the $L(\mathrm{H}\beta) \text{-} \sigma$ correlation, 
still need to be better understood. These consist of the size of the starburst, the age of the burst, the oxygen 
abundance of HII galaxies and the internal extinction correction \citep{2016MNRAS.462.2431C}. The scatter found in 
this $L(\mathrm{H}\beta) \text{-} \sigma$ relation indicates that it probably depends on a second parameter. Some 
progress has been made to mitigate these uncertainties. For example, \citet{2016MNRAS.462.2431C} found that, for 
samples of local HII galaxies, the size of the star forming region can be used as this second parameter. Another 
important consideration is the exclusion of systems supported by rotation, which obviously distorts the 
$L(\mathrm{H}\beta) \text{-} \sigma$ relationship. \citet{1988MNRAS.235..297M} and \citet{2016MNRAS.462.2431C} suggested 
using an upper limit of the velocity dispersion $\log \big[ \sigma (\mathrm{H} \beta) / \mathrm{km}\,\mathrm{s}^{-1} 
\big] \sim 1.8$ to minimize this possibility, although the catalog of suitable sources is then greatly reduced. 
However, even with this limitation, there is no guarantee that this systematic effect is completely eliminated.

\subsection{Geometric relation in the FLRW universe}
In the FLRW metric, the radial comoving distance $d_c (z)$ of a galaxy at redshift $z$ is expressed as 
\begin{align}
    d_c (z) = c \int_0^z \frac{\rd z'}{H(z')} \;. \label{equ:d_C}
\end{align}
The relation between comoving and luminosity distances changes as the sign of $\Omega_{k}$ changes: 
\begin{align}
    \frac{d_L (z)}{1+z} = \begin{cases}
        \displaystyle\frac{c}{H_0} \frac{1}{\sqrt{\Ok}} \sinh \left[ \sqrt{\Ok} \frac{H_0}{c} \, 
        d_c (z)  \right] \quad &\text{for} \;\Ok > 0 \\
        d_c (z) \quad &\text{for} \;\Ok = 0 \\
        \displaystyle\frac{c}{H_0} \frac{1}{\sqrt{|\Ok|}} \sin \left[ \sqrt{|\Ok|} \frac{H_0}{c} \, 
        d_c (z)  \right] \quad &\text{for} \;\Ok < 0 \\
    \end{cases} \;. \label{equ:dL_dC}
\end{align}

To find a relation between the HIIGx and $H(z)$ data, we note that the derivative of Equation~\eqref{equ:d_C} simply gives $d^\prime_c (z) \equiv d/dz(d_c[z]) = c / H(z)$. This suggests a similarly useful operation with Equation~\eqref{equ:dL_dC}, from which a new relation between $H(z)$ and the luminosity distance may be extracted:
\begin{align}
    E (z) &\equiv \frac{H (z)}{H_0} \notag \\
    &= \begin{cases}
        \displaystyle \frac{c}{H_0\,f(z)} \sqrt{ 1 + \left[   \frac{H_0\, d_L(z)}{c} \frac{\sqrt{\Ok}}
        {1+z} \right]^2 } \quad &\Ok > 0 \\
        \displaystyle {c} / {\big[ H_0\,f(z) \big]} \quad &\Ok = 0 \\ 
        \displaystyle \frac{c}{H_0\,f(z)} \sqrt{ 1 - \left[  \frac{H_0\, d_L(z)}{c} \frac{\sqrt{|\Ok|} }
        {1+z} \right]^2 } \quad &\Ok < 0
    \end{cases} \;, \label{equ:jkl} \\
\intertext{and}
    f(z) &\equiv \frac{d}{d z} \left[ \frac{d_L(z)}{1+z} \right] = \frac{d'_L(z)}{1+z} - \frac{d_L(z)}{(1+z)^2} \ . 
\end{align}

Equation~\eqref{equ:jkl} relates the luminosity distance $d_L(z)$ to the Hubble expansion rate $H(z)$ in the FLRW universe, in which the former may be extracted with GP reconstruction of the HIIGx Hubble diagram, while the latter may be found using cosmic chronometers. We employ two distinct values of the Hubble constant to turn the Hubble parameter $H(z)$ into a dimensionless quantity. These are the {\it Planck} value and that measured locally using the distance ladder:
\begin{align}
    \text{\textit{Planck} value:}\; \ H_0 &= 67.66 \pm 0.42 \; \mathrm{km}\;\mathrm{s}^{-1}\;
    \mathrm{Mpc}^{-1} \;, \label{equ:H0Planck} \\
    \text{local value:}\,\quad \ H_0 &= 73.24 \pm 1.74 \; \mathrm{km}\;\mathrm{s}^{-1}\;
    \mathrm{Mpc}^{-1} \;. \label{equ:H0local}
\end{align}
For each of these quantitites, we extract a purely geometric measurement of the curvature parameter $\Omega_{k}$ though, as noted earlier, our intention is clearly to probe which of these two disparate values of $H_0$ is more consistent with spatial flatness.

For consistency with the GP results, we represent the dimensionless luminosity distance $(H_0\, d_L)/c$ using the (shifted) distance modulus $\eta$ (Equation \eqref{equ:nudef}). The dimensionless Hubble parameter is 
\begin{align}
    E (z; \Omega_{k}) &= \begin{cases}
        \displaystyle g(z) \sqrt{ 1 + \left[  \frac{\sqrt{\Ok}  \, 10^{\eta(z)/5}}{ { \big[ c/ (\mathrm{km}\,\mathrm{s}^{-1}) \big]} (1+z)} \right]^2 } \quad &\Ok > 0 \\
        \displaystyle  g(z) \quad &\Ok = 0 \\ 
        \displaystyle g(z) \sqrt{ 1 - \left[   \frac{\sqrt{|\Ok|} \, 10^{\eta(z)/5}}{{ \big[ c/ (\mathrm{km}\,\mathrm{s}^{-1}) \big]} (1+z)} \right]^2 } \quad &\Ok < 0
    \end{cases} \;, \\
\intertext{where}
    g(z) &= \bigg[ \frac{1}{5z(1+z)} \frac{10^{\eta(z)/5} \, {\eta}^\prime(z)}{{  c/ (\mathrm{km}\,\mathrm{s}^{-1}) }} -    \frac{1}{(1+z)^2} \frac{10^{\eta(z)/5}}{{  c/ (\mathrm{km}\,\mathrm{s}^{-1}) }} \bigg]^{-1} \ . \label{equ:gz}
\end{align}

\subsection{Uncertainty of $E (z; \Omega_k)$}
The covariance between $\eta$ and $\eta^\prime$ may be found with GP reconstruction, i.e., $\Cov \big(\eta[z], \eta^\prime[z] \big) \neq 0$.  In the context of GP, the values $\eta_\star \equiv \eta(z_\star)$ and $\eta^\prime_\star \equiv \eta^\prime (z_\star)$ at any redshift point $z_\star$ follow a multivariate Gaussian distribution \citep{2012JCAP...06..036S}
\begin{align}
    \begin{pmatrix}
        \eta_\star \\ \eta^\prime_\star
    \end{pmatrix} \sim \mathcal{N} \left[ \begin{pmatrix}
        \bar{\eta}_\star \\ \bar{\eta^\prime}_\star
    \end{pmatrix}, \begin{pmatrix}
        \Var (\eta_\star) & \Cov (\eta_\star, \eta^\prime_\star) \\
        \Cov (\eta_\star, \eta^\prime_\star) & \Var (\eta^\prime_\star)
    \end{pmatrix}  \right]\ ,  \label{equ:mdmdm}
\end{align}
where $\bar{\eta}$, $\bar{\eta^\prime}, \Cov(\eta, \eta^\prime)$ and $\Var (\eta, \eta^\prime)$ are computed with the GP code called GaPP\footnote{\url{http://www.acgc.uct.ac.za/~seikel/GAPP/main.html}} developed by  \citet{2012JCAP...06..036S}.

For every redshift point $z_i$ at which a measurement of $H(z_i)$ is made, the uncertainty $\sigma_{E_i(\Omega_k)}$ is determined by Monte Carlo sampling, where $\begin{pmatrix} \eta_i \\ \eta^\prime_i \end{pmatrix}$ is extracted from the probability distribution given by Equation \eqref{equ:mdmdm}. For example, we generate $N_{\mathrm{MC}} = 10^4$ points,
\begin{align}
    \left\{ \eta_i^{(j)}, {\eta^\prime_i}^{(j)} \right\}_{j=1}^{N_{\mathrm{MC}} } \quad \Rightarrow \quad 
    \left\{ E_i\!\left(\eta_i^{(j)}, {\eta^\prime_i}^{(j)} \right) \right\}_{j=1}^{N_{\mathrm{MC}} } \ ,
\end{align}
and from these calculate the standard deviation $\sigma_{E_i}$ of this array.

\section{Data}
\label{sec:data}

\subsection{Hubble parameter from cosmic chronometers}
\begin{figure}
    \plotone{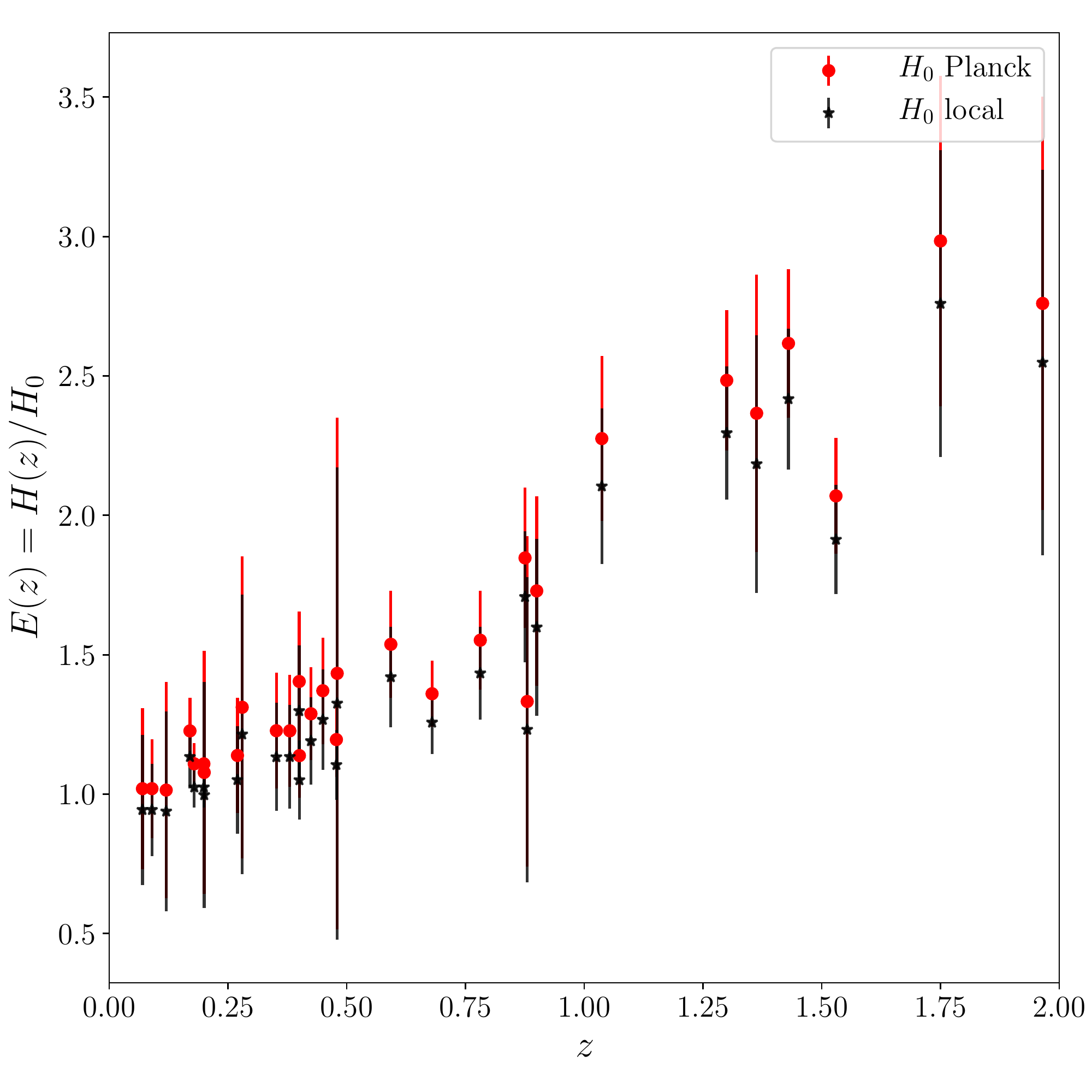}
    \caption{Dimensionless Hubble parameter $E(z) \equiv H(z) / H_0$ data, using two distinct values of the Hubble constant $H_0$, one from \textit{Planck} (red circle) and the other from the local distance ladder (black star), provided in Equations \eqref{equ:H0Planck} and \eqref{equ:H0local}. The uncertainties in $E(z)$ are due to the uncertainties of both $H(z)$ and $H_0$, which are estimated by a Monte Carlo method. All of these data are based on observations of cosmic chronometers, listed in Table~\ref{tab:hz}. \label{fig:Ez}}
\end{figure}

The Hubble parameter, $H(z) \equiv \dot{a}/a$, is the expansion rate of the FLRW universe in terms of the scale factor $a(t)$ and its time derivative $\dot{a} \equiv \rd a/ \rd t$. The Hubble expansion rate may be deduced in a model independent fashion from cosmic chronometers, using the differential ages of galaxies, as proposed by \citet{2002ApJ...573...37J}.  We use a sample of 30 $H(z)$ measurements in the redshift range of $0.07 < z < 2.0$, compiled by \citet{2016JCAP...05..014M}, which we list in Table \ref{tab:hz}. The corresponding dimensionless values in this sample are plotted in Figure~\ref{fig:Ez}, using the two distinct values of $H_0$ shown in Equations~\eqref{equ:H0Planck} and \eqref{equ:H0local}.

\begin{table}
    \caption{Hubble Parameter $H(z)$ from Cosmic Chronometers}
    \label{tab:hz}
\begin{tabular}{lcl} 
    \hline
    $z$ & $H(z) \ (\mathrm{km}\,\mathrm{s}^{-1}\,\mathrm{Mpc}^{-1}) $ & References \\
    \hline 
    0.09 & 69  $\pm$  12   & \citet{2003ApJ...593..622J} \\
    \hline
    0.17 & 83  $\pm$  8    & \citet{2005PhRvD..71l3001S} \\
    0.27 & 77  $\pm$  14   & \\
    0.4  & 95 $\pm$ 17     & \\
    0.9  &117 $\pm$ 23     & \\
    1.3  &168 $\pm$ 17     & \\
    1.43 &177 $\pm$ 18     & \\
    1.53 &140 $\pm$ 14     & \\ 
    1.75 &202 $\pm$ 40     & \\
    \hline
    0.48 & 97  $\pm$  62   & \citet{2010JCAP...02..008S} \\
    0.88 & 90  $\pm$  40   & \\
    \hline  
    0.1791 &  75  $\pm$  4 & \citet{2012JCAP...07..053M}\\ 
    0.1993 & 75  $\pm$  5  & \\
    0.3519 & 83  $\pm$  14 & \\
    0.5929 & 104  $\pm$  13 & \\
    0.6797 & 92  $\pm$  8  & \\
    0.7812 & 105  $\pm$  12 & \\
    0.8754 & 125  $\pm$  17 & \\
    1.037  & 154  $\pm$  20 & \\
    \hline   
    0.07   & 69  $\pm$  19.6   & \citet{2014RAA....14.1221Z} \\
    0.12   & 68.6  $\pm$  26.2 & \\  
    0.2    & 72.9  $\pm$  29.6 & \\
    0.28   & 88.8  $\pm$  36.6 & \\ 
    \hline 
    1.363  & 160  $\pm$  33.6  & \citet{2015MNRAS.450L..16M}  \\
    1.965  & 186.5  $\pm$  50.4 & \\ 
    \hline 
    0.3802 & 83  $\pm$  13.5   & \cite{2016JCAP...05..014M} \\
    0.4004 & 77  $\pm$  10.2   & \\
    0.4247 & 87.1  $\pm$  11.2 & \\
    0.4497 & 92.8  $\pm$  12.9 & \\
    0.4783 & 80.9  $\pm$  9    & \\
\hline 
\end{tabular}
\end{table}

\begin{figure}
    \plotone{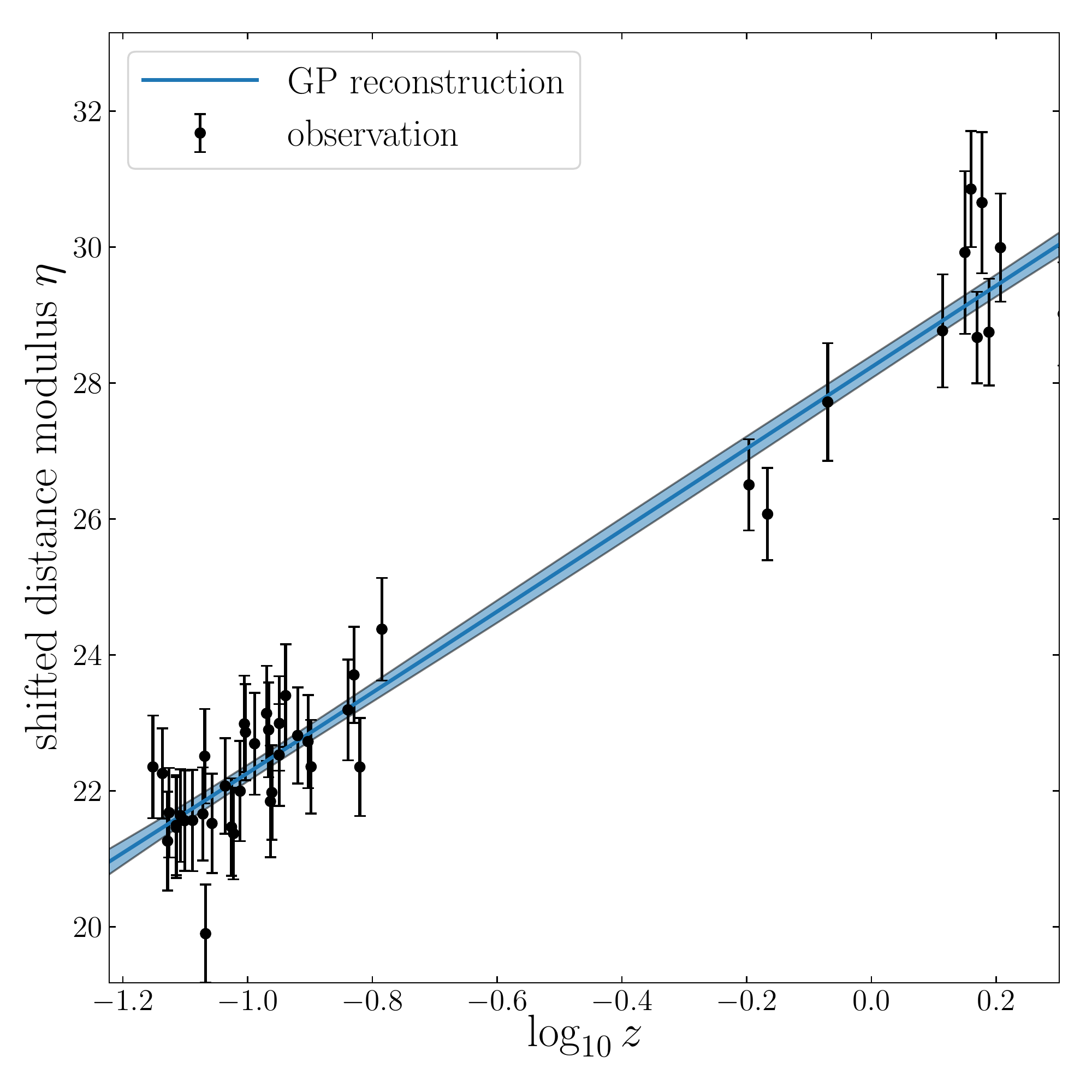}
    \caption{(Shifted) Distance modulus of the currently available HIIGx observations, shown with $1\sigma$ error bars, spanning a redshift range $0.07\lesssim z\lesssim 2.33$. The GP reconstructed (shifted) distance modulus $\eta(z)$ is shown as a solid blue curve, with its $1\sigma$ confidence region (the swath with a lighter shade of blue). The sample consists of 25 high-$z$ HIIGx, 24 giant extragalactic HII regions and and 10 local HIIGx with $z>0.07$ (\citet{2016MNRAS.463.1144W}, see also \citet{terlevich2015road}). \label{fig:HIIGx_rec}} 
\end{figure}

\subsection{Luminosity Distance from GP Reconstruction of the HII Galaxy Hubble Diagram}
For the HIIGx Hubble diagram, we extract the 25 high-$z$ HIIGx, 24 giant extragalactic HII regions and 10 local HIIGx (with $z>0.07$) from the catalog compiled by \citet{2016MNRAS.463.1144W}, from the observational works of \citet{2005ApJ...635L..21H},
\citet{2006ApJ...646..107E}, \citet{2014ApJ...791...17M}, \citet{2014ApJ...785..153M}, \citet{doi:10.1093/mnras/stu987}
and \citet{terlevich2015road}. We exclude other local HIIGx in this catalog because most of them (i.e., $97/107$) have a 
redshift less than the minimum redshift ($z_{\mathrm{min}} = 0.07$) sampled in the cosmic-chronometer data. The GP reconstructed `shifted' distance modulus, $\eta\Big\{z,\sigma_{\eta}(z),\eta^\prime(z),\sigma_{\eta^\prime}(z), \Cov \big[\eta(z),\eta^\prime(z) \big] \Big\}$, is calculated from these data, following the prescription  described in \citet{yennapureddy2017reconstruction}. The distance modulus data and GP reconstruction are shown together in Figure~\ref{fig:HIIGx_rec}. As one may see in this figure, the error of the reconstructed $\eta(z)$ function is smaller than that of the original HIIGx data.
The error calculated by the GP reconstruction depends on the errors of observational data $\sigma_{\eta_{\mathrm{obs}}}$, on the optimized hyperparameter(s) of GP method, such as the characteristic `bumpiness' parameter $\sigma_f$, and on the product of the covariance matrixes $K_* K^{-1} K_*^T$ between the estimation point $z_*$ and dataset points $\{ z_i \}$ (see \citet{2012JCAP...06..036S,yennapureddy2017reconstruction}). 
The reconstruction uncertainty $\sigma_{\eta_{\mathrm{GP}}} (z_*)$ at $z_*$ will be smaller than $\sigma_{\eta_{\mathrm{obs}}}$ when there is a large correlation between the data for the point $z_*$: $K_* K^{-1} K_*^T > \sigma_f$, which is the most common case with the HII-galaxy data used in this study. Thus the estimated confidence region is smaller than that of the observational data.

\section{Results and Discussion}
\label{sec:res}

An optimized value of $\Omega_k$ may be extracted in a model-independent fashion from fitting the 30 $E^{\mathrm{cc}}\{z_i, \sigma_{E^{\mathrm{cc}}, i} \}$ cosmic-chronometer and $E\{z_i, \sigma_{E, i} \}$ GP-reconstructed values of the dimenionless Hubble constant. We use Bayesian statistical methods and the Markov Chain Monte Carlo (MCMC) technique to calculate the posterior probability density function (PDF) of $\Omega_k$, given as
\begin{align}
    p(\Omega_k | \mathrm{data}) \propto \mathcal{L} (\Omega_k, \mathrm{data}) 
    \times p_{\mathrm{prior}} (\Omega_k) \ ,
\end{align}
where 
\begin{enumerate}
    \item $\mathcal{L} (\Omega_k, \mathrm{data}) \propto \exp (-\chi^2/2)$ is the likelihood function, and  
    \begin{align}
        \chi^2 (\Omega_k) = \sum_{i=1}^{N=30} \left\{ \frac{\big[E^{\mathrm{cc}}_i - E_i (\Omega_k)\big]^2 }{\sigma^2_{E^{\mathrm{cc}}, i} + \sigma^2_{E^{\mathrm{L}}, i} (\Omega_k) } \right\}
    \end{align}
is the chi-squared function;
    \item $p_{\mathrm{prior}} (\Omega_k)$ is the prior of $\Omega_k$ and (assumed) uniform 
distribution between $-1$ and 1.    
\end{enumerate}

\begin{figure}
    \plotone{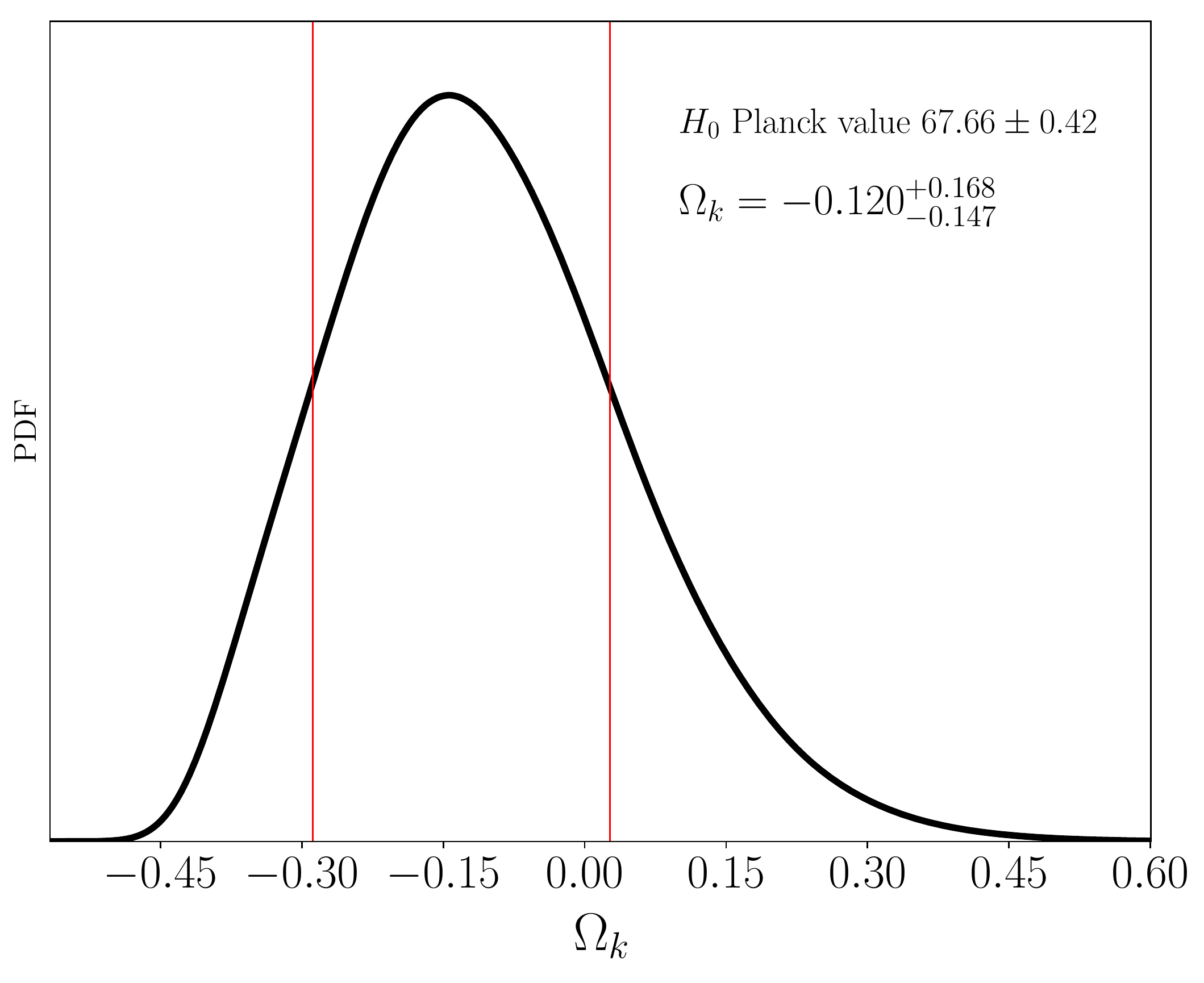}
    \caption{Posterior probability density function of the parameter $\Omega_k$, for the Hubble constant measured by \textit{Planck}, i.e., $H_0=67.66 \pm 0.42\,\mathrm{km}\,\mathrm{s}^{-1}\,\mathrm{Mpc}^{-1}$, showing also the optimized value and its $1\sigma$ error. \label{fig:Okpdf_Planck}}
\end{figure}

\begin{figure}
    \plotone{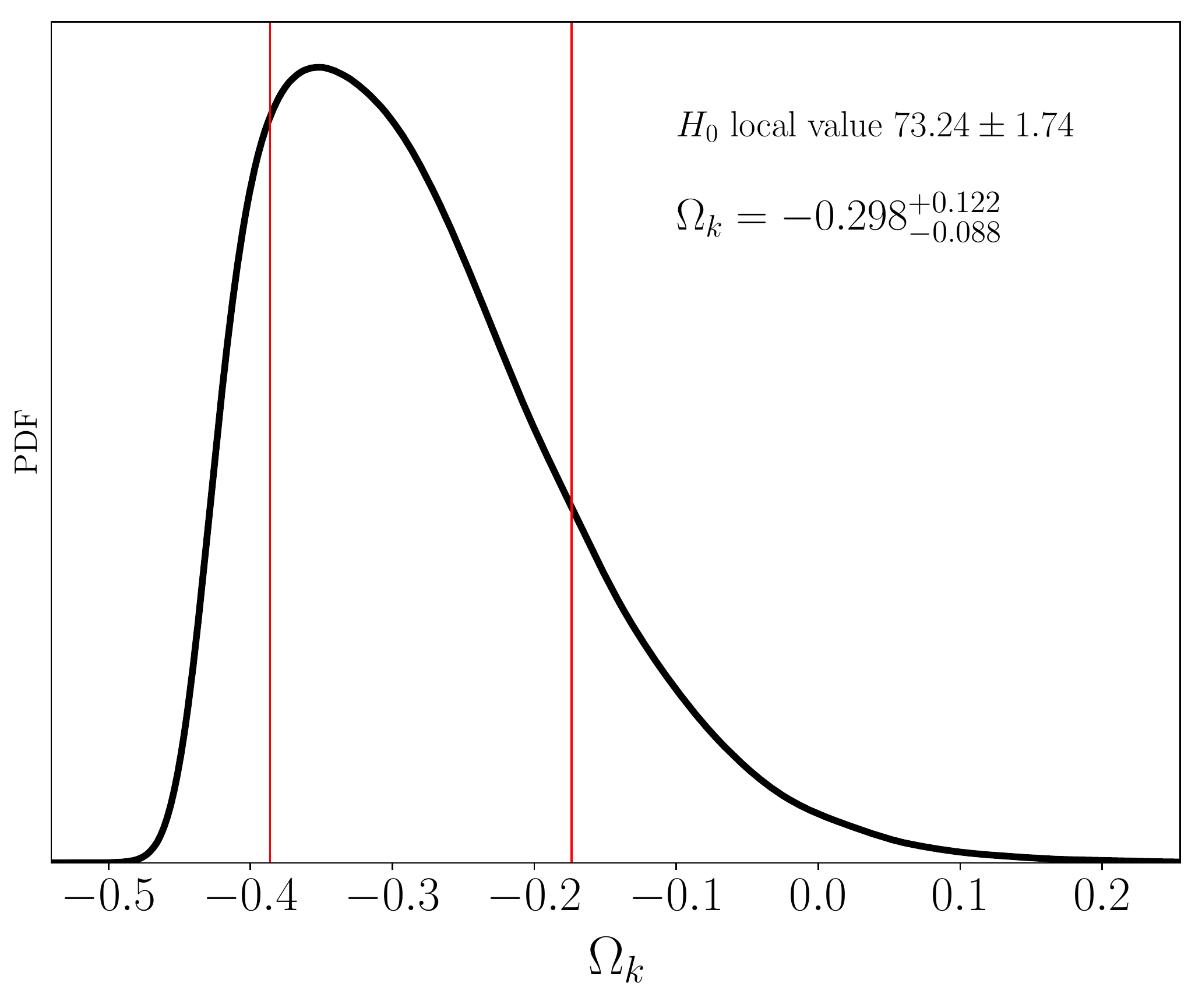}
    \caption{Same as Figure~\ref{fig:Okpdf_Planck}, but for the Hubble constant measured with the local distance ladder, i.e. $H_0=73.24 \pm 1.74 \,\mathrm{km}\,\mathrm{s}^{-1}\,\mathrm{Mpc}^{-1}$, showing also the optimized value and its $1\sigma$ error. \label{fig:Okpdf_local}}
\end{figure}

We use the Python module emcee\footnote{\url{http://dfm.io/emcee/current/}} \citep{2013PASP..125..306F} to sample from the posterior distribution of $\Omega_k$, and find that its optimized value and $1\sigma$ error are
\begin{align}
    \Omega_k &= -0.120^{+0.168}_{-0.147} \quad (H_0\ \mathrm{Planck\ value}) \ , \\
    \Omega_k &= -0.298^{+0.122}_{-0.088} \quad (H_0\ \mathrm{local\ distance-ladder\ value}) \ ,
\end{align}
for the two distinct values of $H_0$. The two corresponding PDF plots are shown in Figures~\ref{fig:Okpdf_Planck} and \ref{fig:Okpdf_local}, respectively.

We see that the {\it Planck} measured value of $H_0$ (Figure~\ref{fig:Okpdf_Planck}) is consistent with spatial flatness to within $1\sigma$. This is quite meaningful in the sense that the {\it Planck} optimization procedure is based on the analysis of anisotropies in the CMB, thought to have originated as quantum fluctuations in the inflaton field. Self-consistency would demand that the value of $\Omega_k$ calculated with the {\it Planck} Hubble constant should therefore support a spatially flat Universe---an unavoidable consequence of the inflationary paradigm.

In contrast, we also see that the Hubble constant measured with the local distance ladder (Figure~\ref{fig:Okpdf_local}) is in significant tension with the requirements of a flat Universe. Our results show that $\Omega_k$ measured in this way rules out spatial flatness at roughly $3\sigma$. Thus, if this locally measured Hubble constant is a true reflection of the large-scale cosmic expansion rate, our results could be taken as some evidence {\it against} inflation as the true solution to the horizon problem (see, e.g., \citealt{2013A&A...553A..76M,2018EPJC...78..739M}).

The disparity between the two distinct values of $H_0$ may in fact be real, signaling the role of local physics in changing the nearby expansion rate compared to what we see on the largest cosmic scales. Some authors have speculated on the possibility that a local ``Hubble bubble" \citep{1997ApJ...486...32S,2013ApJ...775...62K,2015EL....10939002E,2017arXiv170609734C,Romano:2016utn} might be influencing the local dynamics within a distance $\sim 300$ Mpc (i.e., $z\lesssim 0.07$). If true, such a fluctuation might lead to anomalous velocities within this region, causing the nearby expansion to deviate somewhat from a pure Hubble flow. This effect could be the reason we are seeing nearby velocities slightly larger than Hubble, implying larger than expected luminosity distances at redshifts smaller than $\sim 0.07$. Our findings would be fully consistent with this scenario, given that the data we have used in this paper pertain to sources at redshifts well outside the so-called Hubble bubble. We would therefore expect our analysis to support the {\it Planck} value of $H_0$, rather than the locally measured one. 

\begin{figure}
    \plotone{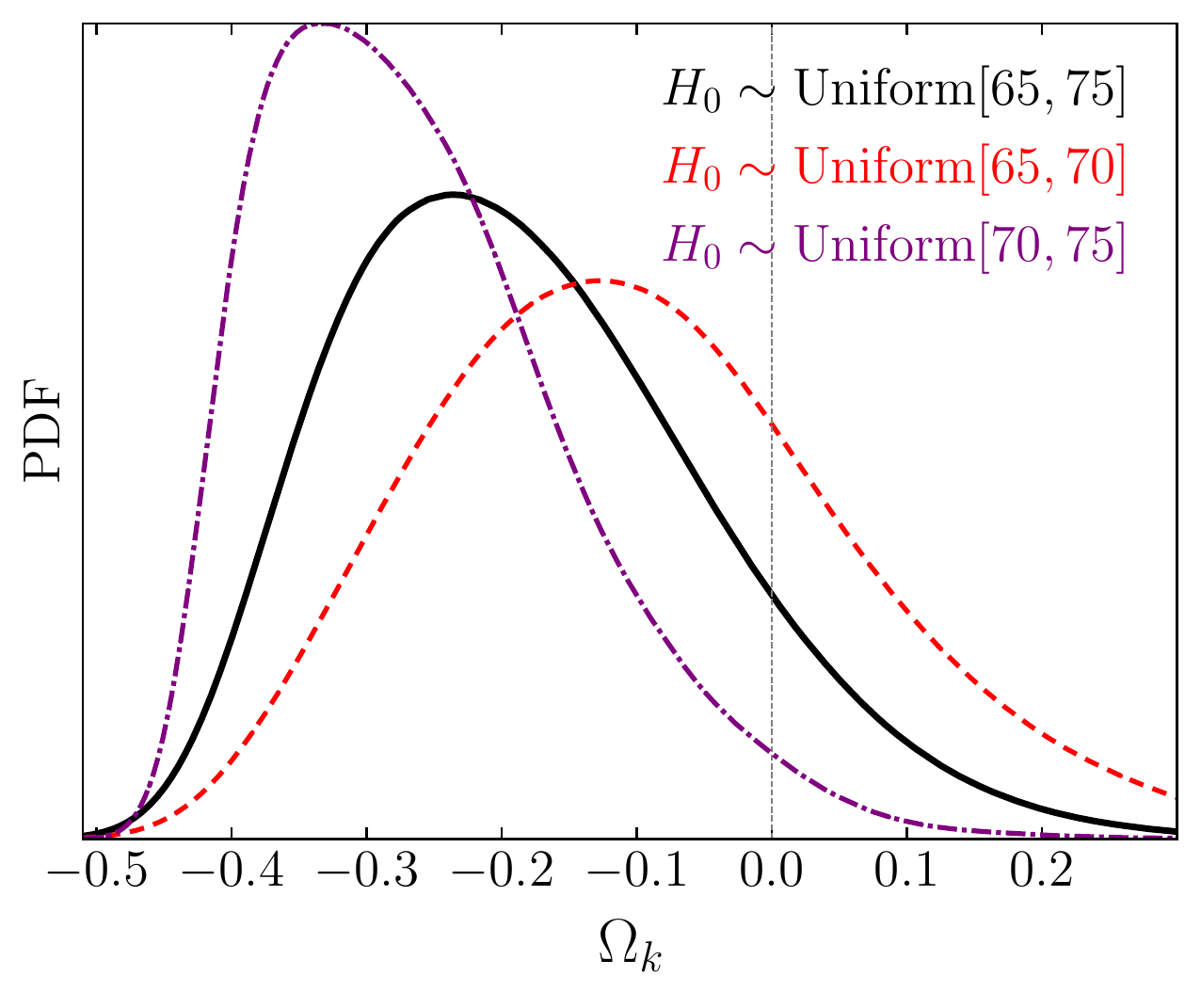}
    \caption{Same as Figure~\ref{fig:Okpdf_Planck}, except for a Hubble constant with a uniform distribution 
in the intervals $[65, 75], [65, 70]$ and $[70, 75]$ $\mathrm{km}\,\mathrm{s}^{-1} \,\mathrm{Mpc}^{-1}$, 
respectively. \label{fig:Okpdf_flat}}
\end{figure}

To further validate our results, we have also optimized the value of $\Omega_k$ corresponding to an $H_0$ with a flat 
probability distribution instead of Gaussian. Considering that the typical $H_0$ \textit{Planck} and local 
distance-ladder values are approximately $67$ and $73$ $\mathrm{km}\,\mathrm{s}^{-1} \,\mathrm{Mpc}^{-1}$, respectively, 
we have chosen uniform distributions over three intervals: $[65, 75], [65, 70]$ and $[70, 75]$ (hereafter all values of 
$H_0$ are presented in terms of $\mathrm{km}\, \mathrm{s}^{-1} \, \mathrm{Mpc}^{-1}$ for conciseness). Following the 
same data handling pipeline as before, we find that the optimized values and their 1-$\sigma$ errors are 
\begin{align}
    \Omega_k &= -0.199^{+0.157}_{-0.126} \ , \ H_0 \sim \mathrm{U} [65, 75] \ , \\
    \Omega_k &= -0.110^{+0.171}_{-0.151} \ , \ H_0 \sim \mathrm{U} [65, 70] \ , \\
    \Omega_k &= -0.279^{+0.129}_{-0.098} \ , \ H_0 \sim \mathrm{U} [70, 75] \ , 
\end{align}
respectively. The corresponding posterior probability distributions are shown in Figure~\ref{fig:Okpdf_flat}.

These curvature constraints validate our conclusion regarding the consistency of $H_0$ with spatial flatness.
The lower $H_0$ value uniformly distributed between $65$ and $70$ is in accordance with $\Omega_k = 0$ to
within 1-$\sigma$, while the larger one is not. Therefore, non-informative distributions of $H_0$ have little 
impact on our curvature fitting results.

\section{Summary}
\label{sec:conc}
In this paper, we have presented a novel approach to the measurement of the spatial flatness parameter, proportional to $\Omega_k$, which avoids possible biases introduced with the pre-adoption of a particular cosmological model. Our first application of this method, reported here, has already yielded a significant result, supporting arguments in favour of the {\it Planck} optimized value of the Hubble constant as being a fair representation of the large-scale cosmic expansion rate. Indeed, the locally measured value of $H_0$ has been ruled out as a true measure of the `average' Hubble constant at a confidence level of $\sim 3\sigma$. 

In this view, our results might also be taken as some evidence in support of the ``Hubble bubble'' concept, which suggests that locally measured expansion velocities somewhat exceed the Hubble flow due to a below average density, thereby implying larger than normal luminosity distances. Quite tellingly, the data we have used are restricted to redshifts $z\gtrsim 0.07$, which also happens to be near the bubble's termination radius. At the very least, all of these inferences are consistent with each other. A stronger case for these conclusions could be made with measurements of the Hubble parameter at redshifts $z\lesssim 0.07$. We shall initiate this investigation in the near future and report the results elsewhere. 

\section*{Acknowledgements}
We are grateful to the anonymous referee for several suggestions
that have led to an improvement in the presentation of our results. This work 
was supported by National Key R \& D Program of China (2017YFA0402600), the National Science Foundation of China 
(Grants No.11573006, 11528306), the Fundamental Research Funds for the Central Universities，and the Special Program for Applied Research on Super Computation of the NSFC-Guangdong Joint Fund (the second phase). 




\bibliography{HIIGx_OHD_curv_ref}



\end{document}